\documentclass[submission]{eptcs}
\providecommand{\event}{TTC 2011}

\usepackage{prelude}

\newcommand{\GROOVE} {{\sc groove}\xspace}

\newcommand{\src} {\emph{src}\xspace}
\newcommand{\trg} {\emph{trg}\xspace}

\newcommand{\edges} {\emph{edges}\xspace}
\newcommand{\nodes} {\emph{nodes}\xspace}
\newcommand{\gcs} {\emph{gcs}\xspace}

\title{Saying Hello World with GROOVE - A Solution to the TTC 2011 Instructive Case}

\author{Amir Hossein Ghamarian \quad
Maarten de Mol \qquad Arend Rensink \quad Eduardo Zambon
\institute{%
Department of Computer Science\\%
University of Twente, The Netherlands%
}%
\email{\{a.h.ghamarian, mj.demol, rensink, zambon\}@ewi.utwente.nl }
}
\def\titlerunning{Saying Hello World with GROOVE - A Solution to the TTC 2011 Instructive Case}
\def\authorrunning{A.H. Ghamarian, M. de Mol,
A. Rensink \& E. Zambon}
\begin{document}
\maketitle
\begin{abstract}
This report presents a solution to the Hello World case study \cite{helloworldcase}
using \GROOVE. We provide and explain the grammar that we used to solve
the case study. Every requested question of the case study was
solved by a single rule application.
\end{abstract}
%
\section{GROOVE}
\seclabel{groove}
\GROOVE\footnote{Available at \url{http://groove.cs.utwente.nl}} \cite{sttt}
is a general purpose graph transformation tool set that uses simple labelled
graphs. The core functionality of \GROOVE is to recursively apply all rules from
a predefined set (the graph production system -- GPS) to a given start graph,
and to all graphs generated by such applications. This results in a \emph{state
space} consisting of the generated graphs.

The main component of the \GROOVE tool set is the Simulator, a graphical tool
that integrates (among others) the functionalities of rule and host graph
editing, and of interactive or automatic state space exploration.

\subsection{Host Graphs}

In \GROOVE, the host graphs, i.e., the graphs to be transformed, are simple
graphs with nodes and directed labelled edges. In simple graphs, edges do not
have an identity, and therefore parallel edges (i.e., edges with same label, and
source and target nodes) are not allowed. Also, for the same reason, edges may
not have attributes.

In the graphical representation, nodes are depicted as rectangles and edges as
binary arrows between two nodes. Node labels can be either node types or flags.
Node types [resp. flags] are displayed in {\bf bold} [resp. {\it italic}] inside
a node rectangle.

\subsection{Rules}
The transformation rules in \GROOVE are represented by a single graph and
colours and shapes are used to distinguish different elements.
\figref{example-rule} shows a small example rule.
\begin{itemize}
\item {\bf Readers.} The black (continuous thin) nodes and edges must be present
in the host graph for the rule to be applicable and are preserved by the rule
application;
\item {\bf Embargoes.} The red (dashed fat) nodes and edges must be absent in
the host graph for the rule to be applicable;
\item {\bf Erasers.} The blue (dashed thin) nodes and edges must be present in
the host graph for the rule to be applicable and are deleted by the rule
application;
\item {\bf Creators.} The green (continuous fat) nodes and edges are created by
the rule application.
\end{itemize}
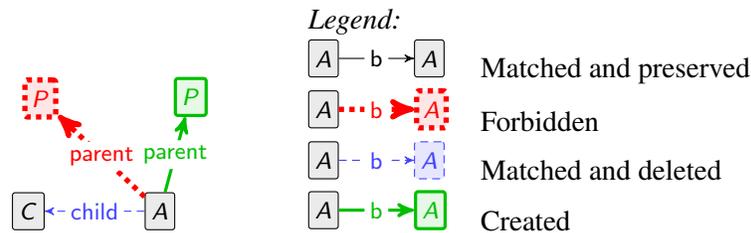
\begin{figure}
\vspace*{5mm}
\centering
\scalebox{\mytikzscale}{\begin{tabular}[t]{@{}l@{}}
\begin{tikzpicture}[scale=\tikzscale]
\node[node] (n3)  at (1.400, -1.265) {\ml{\textit{C}}};
\node[nacnode] (n2)  at (1.480, -0.515) {\ml{\textit{P}}};
\node[newnode] (n1)  at (2.480, -0.505) {\ml{\textit{P}}};
\node[node] (n0)  at (2.275, -1.265) {\ml{\textit{A}}};
\path[newedge] (n0)  -- node[newlab]{parent} (n1) ;
\path[nacedge] (n0)  -- node[naclab]{parent} (n2) ;
\path[deledge](n0.west |- 1.400, -1.265) -- node[dellab]{child} (n3) ;
\userdefinedmacro
\end{tikzpicture}
\renewcommand{\userdefinedmacro}{\relax}\figlabel{example-rule}\end{tabular}}%
 \qquad
\begin{tabular}[b]{ll}
\multicolumn{2}{l}{\it Legend:} \\
\scalebox{\mytikzscale}{\begin{tikzpicture}[scale=\tikzscale]
  \node[node] (n0) {\ml{\textit{A}}};
  \node[node] (n1) [right=of n0] {\ml{\textit{A}}};
  \path (n0) edge[edge] node[lab] {b} (n1);
\end{tikzpicture}} & Matched and preserved \\
\scalebox{\mytikzscale}{\begin{tikzpicture}[scale=\tikzscale]
  \node[node] (n0) {\ml{\textit{A}}};
  \node[nacnode] (n1) [right=of n0] {\ml{\textit{A}}};
  \path (n0) edge[nacedge] node[naclab] {b} (n1);
\end{tikzpicture}} & Forbidden \\
\scalebox{\mytikzscale}{\begin{tikzpicture}[scale=\tikzscale]
  \node[node] (n0) {\ml{\textit{A}}};
  \node[delnode] (n1) [right=of n0] {\ml{\textit{A}}};
  \path (n0) edge[deledge] node[dellab] {b} (n1);
\end{tikzpicture}} & Matched and deleted \\
\scalebox{\mytikzscale}{\begin{tikzpicture}[scale=\tikzscale]
  \node[node] (n0) {\ml{\textit{A}}};
  \node[newnode] (n1) [right=of n0] {\ml{\textit{A}}};
  \path (n0) edge[newedge] node[newlab] {b} (n1);
\end{tikzpicture}} & Created \\
\end{tabular}
\caption{Example \GROOVE rule and legend}
\figlabel{example-rule}
\end{figure}
Embargo elements are usually called Negative Application Conditions (NACs).
When a node type or flag is used in a non-reader element but the node itself is
not modified, the node type or flag is prefixed with a character to indicate its
role. The characters used are $+$, $-$, and $!$, for creator, eraser, and
embargo elements, respectively.

Additional notation and functionalities of the tool are presented along with the
developed solution for the case.

%
%
\section{Solution}
\seclabel{solution}
In this section we describe our solution to the case study.

\subsection{Hello world!}
Metamodels are modelled as type graphs in \GROOVE.
Each node in a type graph corresponds to a node type; some have associated
attributes. Types shown in \textbf{\textit{bold italic}} inside dashed nodes are
abstract. Edges with open triangular arrows indicate type inheritance.

\begin{itemize}
  \item The greeting metamodel of the case study is modelled by the  type graph in \figref{greetingtype}. The rule which makes a node of type greeting is shown in \figref{makegreeting}. The ``text" element of the greeting type is defined as a string attribute.
  \item The type graph presented in \figref{greetingmessagetype} represent the metamodel given in Fig. 2 of the case study.
  The ``text" element of type GreetingMessage and the ``name" element of type Person are modelled using string attributes. \figref{greetingmessage} models the rule that generates the required graph which complies with the type graph.
  \item Every rule in \GROOVE can have a print format property which describes the format in which the rule writes to the standard output when it is applied. The value of attributes with parameters can be written to the standard output by a format similar to that of the C printf. In this case the text attribute of GreetingMessage and the name attribute of Person have parameters 0 and 1 respectively. The print format property of the helloMessage rule is ``The output is \%s \%s \%n",
      in which each ``\%s" refers to the value of a parameter based on its order of appearance in the print format. The application of this rule prints ``The output is Hello TTC Participants" to the standard output.
\end{itemize}

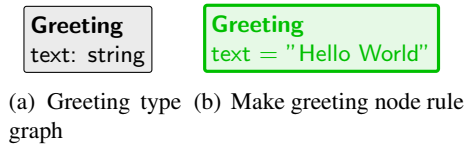
\begin{figure}
\vspace*{5mm}
\centering
\subfigure[Greeting type graph]{
\scalebox{\mytikzscale}{
\scalebox{\mytikzscale}{\begin{tabular}[t]{@{}l@{}}
\begin{tikzpicture}[scale=\tikzscale]
\node[node] (n0)  at (0.845, -0.550) {\ml{\textbf{Greeting}\\text: string}};
\userdefinedmacro
\end{tikzpicture}
\renewcommand{\userdefinedmacro}{\relax}\figlabel{greetingtype}\end{tabular}}%

}
}
\subfigure[Make greeting node rule]{
\scalebox{\mytikzscale}{
\scalebox{\mytikzscale}{\begin{tabular}[t]{@{}l@{}}
\begin{tikzpicture}[scale=\tikzscale]
\node[newnode] (n0)  at (1.100, -0.440) {\ml{\textbf{Greeting}\\text = "Hello World"}};
\userdefinedmacro
\end{tikzpicture}
\renewcommand{\userdefinedmacro}{\relax}\figlabel{makegreeting}\end{tabular}}%

}
}
\caption{Greeting type graph and make greeting node rule}
\end{figure}

\begin{figure}
\centering
\subfigure[Greeting message type graph]{
\scalebox{\mytikzscale}{
\scalebox{\mytikzscale}{\begin{tabular}[t]{@{}l@{}}
\begin{tikzpicture}[scale=\tikzscale]
\node[node] (n0)  at (1.740, -0.745) {\ml{\textbf{Greeting}}};
\node[node] (n2)  at (2.660, -1.700) {\ml{\textbf{Person}\\name: string}};
\node[node] (n1)  at (0.920, -1.690) {\ml{\textbf{GreetingMessage}\\text: string}};
\path[edge] (n0)  -- node[lab]{greetingMessage} (n1) ;
\path[edge] (n0)  -- node[lab]{person} (n2) ;
\userdefinedmacro
\end{tikzpicture}
\renewcommand{\userdefinedmacro}{\relax}\figlabel{greetingmessagetype}\end{tabular}}%

}
}
\subfigure[Make greeting message rule]{
\scalebox{\mytikzscale}{
\scalebox{\mytikzscale}{\begin{tabular}[t]{@{}l@{}}
\begin{tikzpicture}[scale=\tikzscale]
\node[node, attr] (n4)  at (1.170, -1.825) {\ml{"Hello"}};
\node[newnode] (n1)  at (1.220, -1.105) {\ml{\textbf{GreetingMessage}}};
\node[newnode] (n2)  at (3.410, -1.055) {\ml{\textbf{Person}}};
\node[newnode] (n0)  at (2.290, -0.375) {\ml{\textbf{Greeting}}};
\node[node, attr] (n3)  at (3.395, -1.825) {\ml{"TTC Participants"}};
\path[newedge] (n0)  -- node[newlab]{greetingMessage} (n1) ;
\path[newedge](n2.south -| 3.395, -1.825) -- node[newlab]{name} (n3) ;
\path[newedge](n1.south -| 1.170, -1.825) -- node[newlab]{text} (n4) ;
\path[newedge] (n0)  -- node[newlab]{person} (n2) ;
\userdefinedmacro
\end{tikzpicture}
\renewcommand{\userdefinedmacro}{\relax}\figlabel{greetingmessage}\end{tabular}}%

}
}
\caption{Greeting message type graph and rule}
\end{figure}
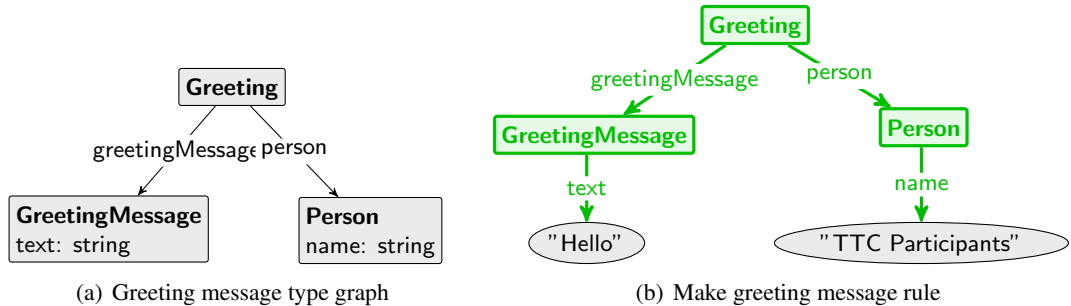

\subsection{Counting questions}
All questions in this section are to compute the number of occurrences of different subgraphs in the context graph.
\GROOVE allows the use of nested universal and existential quantification in the rule description. All occurrences
of subgraph can be captured by a universal quantifier. Each
universal quantifier can report the number of found matches in an ``int" attribute which is connected to
the quantifier by a special edge with the label ``count". Similar to the previous
section, by adding a parameter to the attribute, we can print the number of matches to the standard output.

\begin{figure}
\vspace*{5mm}
\centering
\subfigure[\#nodes]{
\scalebox{\mytikzscale}{
\scalebox{\mytikzscale}{\begin{tabular}[t]{@{}l@{}}
\begin{tikzpicture}[scale=\tikzscale]
\node[node] (n0)  at (2.345, -2.125) {\ml{\textbf{Node}}};
\node[node, attr] (n2)  at (3.375, -2.945) {\ml{\textbf{int}}};
\node[quantnode] (n1)  at (2.320, -2.955) {\ml{$\forall$}};
\path[quantedge](n1.east |- 3.375, -2.945) -- node[lab]{count} (n2) ;
\path[quantedge](n0.south -| 2.320, -2.955) -- node[lab]{at} (n1) ;
\userdefinedmacro
\end{tikzpicture}
\renewcommand{\userdefinedmacro}{\relax}\figlabel{numberofnodes}\end{tabular}}%

}
}
\subfigure[\#looping edges]{
\scalebox{\mytikzscale}{
\scalebox{\mytikzscale}{\begin{tabular}[t]{@{}l@{}}
\begin{tikzpicture}[scale=\tikzscale]
\node[node, attr] (n3)  at (2.405, -1.815) {\ml{\textbf{int}}};
\node[node] (n0)  at (1.970, -0.915) {\ml{\textbf{Edge}}};
\node[quantnode] (n2)  at (1.310, -1.685) {\ml{$\forall$}};
\node[node] (n1)  at (0.665, -0.855) {\ml{\textbf{Node}}};
\path[quantedge](n2.east |- 2.405, -1.815) -- node[lab]{count} (n3) ;
\path[edge](n0.west |- 0.665, -0.855) -- node[lab]{src} (n1) ;
\path[edge] (n0) .. controls (1.680, -0.430) and (1.020, -0.400) ..  (n1) ;
\node[lab] at (1.348, -0.433){trg};
\path[quantedge] (n0)  -- node[lab]{at} (n2) ;
\path[quantedge] (n1)  -- node[lab]{at} (n2) ;
\userdefinedmacro
\end{tikzpicture}
\renewcommand{\userdefinedmacro}{\relax}\figlabel{loopingedges}\end{tabular}}%

}
}
\subfigure[\#isolated nodes]{
\scalebox{\mytikzscale}{
\scalebox{\mytikzscale}{\begin{tabular}[t]{@{}l@{}}
\begin{tikzpicture}[scale=\tikzscale]
\node[quantnode] (n3)  at (1.560, -0.995) {\ml{$\forall$}};
\node[node] (n0)  at (2.595, -0.995) {\ml{\textbf{Node}}};
\node[nacnode] (n2)  at (3.570, -1.805) {\ml{\textbf{Edge}}};
\node[nacnode] (n1)  at (3.590, -0.335) {\ml{\textbf{Edge}}};
\node[node, attr] (n4)  at (1.565, -1.755) {\ml{\textbf{int}}};
\path[quantedge] (n1)  -- node[lab]{at} (n3) ;
\path[quantedge] (n2)  -- node[lab]{at} (n3) ;
\path[nacedge] (n2)  -- node[naclab]{trg} (n0) ;
\path[nacedge] (n1)  -- node[naclab]{src} (n0) ;
\path[quantedge](n3.south -| 1.565, -1.755) -- node[lab]{count} (n4) ;
\path[quantedge](n0.west |- 1.560, -0.995) -- node[lab]{at} (n3) ;
\userdefinedmacro
\end{tikzpicture}
\renewcommand{\userdefinedmacro}{\relax}\figlabel{isolatednodes}\end{tabular}}%

}
}
\subfigure[\#cycles of three]{
\scalebox{\mytikzscale}{
\scalebox{\mytikzscale}{\begin{tabular}[t]{@{}l@{}}
\begin{tikzpicture}[scale=\tikzscale]
\node[node, attr] (n4)  at (5.445, -4.025) {\ml{\textbf{int}}};
\node[node] (n1)  at (4.345, -2.455) {\ml{\textbf{Node}}};
\node[quantnode] (n2)  at (4.410, -3.925) {\ml{$\forall$}};
\node[node] (n0)  at (2.515, -2.485) {\ml{\textbf{Node}}};
\node[node] (n3)  at (2.265, -3.915) {\ml{\textbf{Node}}};
\path[quantedge](n3.east |- 4.410, -3.925) -- node[lab]{at} (n2) ;
\path[quantedge](n2.east |- 5.445, -4.025) -- node[lab]{count} (n4) ;
\path[edge] (n0)  -- node[lab]{\textit{$-$src.trg}} (n3) ;
\path[edge](n1.west |- 2.515, -2.485) -- node[lab]{\textit{$-$src.trg}} (n0) ;
\path[edge, -] (n1) .. controls (3.890, -1.990) and (2.980, -2.000) ..  (n0) ;
\node[lab] at (3.438, -2.001){\textit{!=}};
\path[quantedge] (n0)  -- node[lab]{at} (n2) ;
\path[quantedge](n1.south -| 4.410, -3.925) -- node[lab]{at} (n2) ;
\path[edge, -] (n0) .. controls (2.060, -2.780) and (1.930, -3.500) ..  (n3) ;
\node[lab] at (2.003, -3.146){\textit{!=}};
\path[edge, -] (n1) .. controls (4.110, -3.170) and (3.220, -3.810) ..  (n3) ;
\node[lab] at (3.595, -3.495){\textit{!=}};
\path[edge] (n3)  -- node[lab]{\textit{$-$src.trg}} (n1) ;
\userdefinedmacro
\end{tikzpicture}
\renewcommand{\userdefinedmacro}{\relax}\figlabel{cyclesofthree}\end{tabular}}%

}
}
\subfigure[\#dangling edges]{
\scalebox{\mytikzscale}{
\scalebox{\mytikzscale}{\begin{tabular}[t]{@{}l@{}}
\begin{tikzpicture}[scale=\tikzscale]
\node[quantnode] (n3)  at (3.110, -3.235) {\ml{$\forall$}};
\node[node] (n0)  at (3.110, -2.515) {\ml{\textbf{Edge}}};
\node[nacnode] (n2)  at (4.415, -1.695) {\ml{\textbf{Node}}};
\node[nacnode] (n1)  at (1.865, -1.675) {\ml{\textbf{Node}}};
\node[node, attr] (n4)  at (4.385, -3.295) {\ml{\textbf{int}}};
\path[quantedge] (n1)  -- node[lab]{at} (n3) ;
\path[quantedge] (n2)  -- node[lab]{at} (n3) ;
\path[nacedge, -](n1.east |- 4.415, -1.695) -- node[lab]{$+$} (n2) ;
\path[quantedge](n3.east |- 4.385, -3.295) -- node[lab]{count} (n4) ;
\path[quantedge](n0.south -| 3.110, -3.235) -- node[lab]{at} (n3) ;
\path[nacedge] (n0)  -- node[naclab]{src} (n1) ;
\path[nacedge] (n0)  -- node[naclab]{trg} (n2) ;
\userdefinedmacro
\end{tikzpicture}
\renewcommand{\userdefinedmacro}{\relax}\figlabel{danglingedges}\end{tabular}}%

}
}
\caption{Counting rules}
\end{figure}
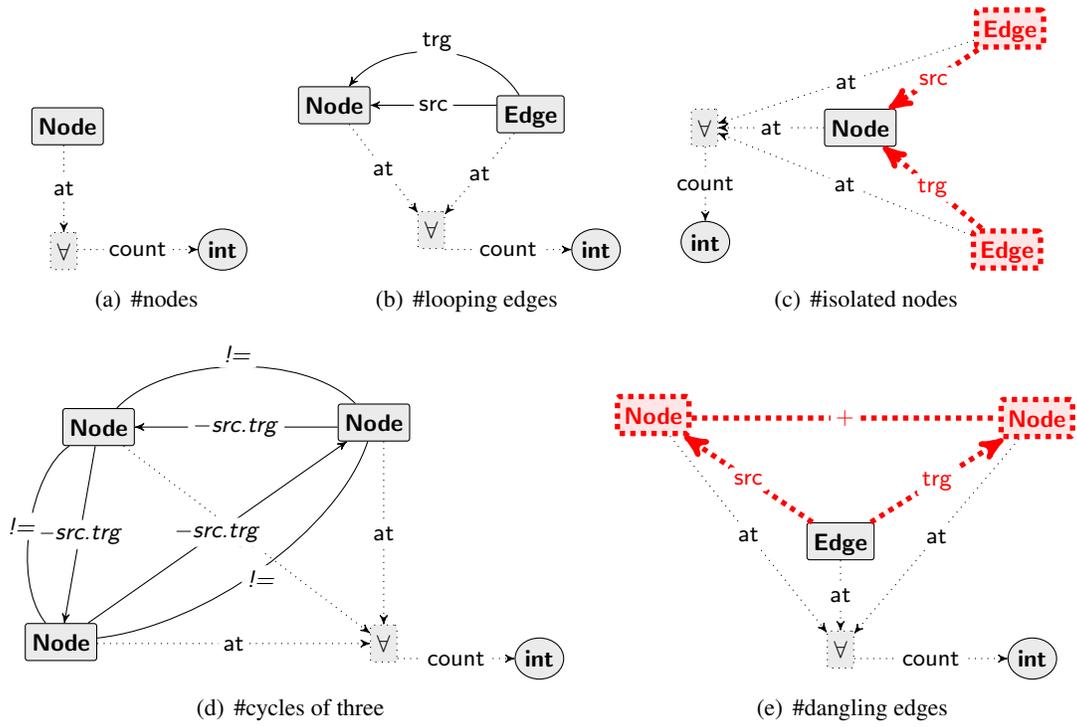

\begin{description}
  \item[\#nodes] As represented in \figref{numberofnodes}, the number of nodes of type Node can be found by a universally quantified node of type Node.

  \item[\#looping edges] Similarly, for finding the loop edges, we only need to universally quantify a looping edge (see \figref{loopingedges}).

  \item[\#isolated nodes] An isolated node is described by a node of type Node with two NACs for both \src and \trg (shown in
  \figref{isolatednodes}). A universal quantifier finds and counts
  all occurrences.

  \item[\#cycles of three] \figref{cyclesofthree} shows the rule for counting the number of cycles with three nodes. A cycle with three nodes
  is trivially described by three connected nodes. The edge with the label ``!=" (injectivity constraint) ensures that the nodes are pairwise different,
  and the edges with label ``-\src.\trg" are abbreviations for paths with two edges, with labels \src and \trg, respectively.
  The first edge with label src must have a reverse direction. The number of occurrences can be counted by universally quantifying this cycle.

 \item[\#dangling edges] This rule is very similar to the isolated nodes case. Dangling edges are nodes with type Edge of which \textbf{at least one} of its outgoing edges (\src or \trg) is missing, however, isolated nodes are nodes of type Node with \textbf{both} incoming \src and \trg edges missing. Therefore, for specifying dangling edges we need to specify a disjunctive relation between the two NACs for the absence of \src and \trg. This disjunctive relation between NACs is specified by an edge with label ``+" in \figref{danglingedges}.
\end{description}

\subsection{Reverse edges}
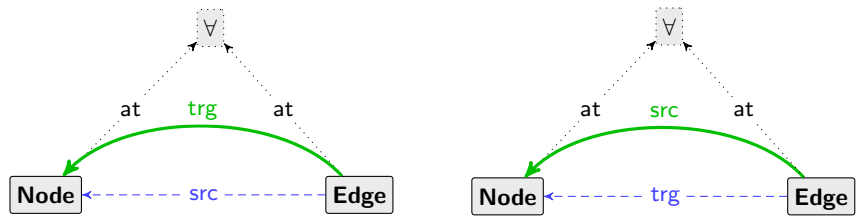
\begin{figure}
\vspace*{5mm}
\centering
\begin{tikzpicture}[scale=\tikzscale]
\node[node] (n4)  at (5.610, -1.865) {\ml{\textbf{Edge}}};
\node[node] (n5)  at (3.525, -1.865) {\ml{\textbf{Node}}};
\node[node] (n1)  at (2.540, -1.855) {\ml{\textbf{Edge}}};
\node[quantnode] (n7)  at (4.600, -0.735) {\ml{$\forall$}};
\node[node] (n0)  at (0.455, -1.855) {\ml{\textbf{Node}}};
\node[quantnode] (n3)  at (1.550, -0.745) {\ml{$\forall$}};
\path[quantedge] (n5)  -- node[lab]{at} (n7) ;
\path[quantedge] (n1)  -- node[lab]{at} (n3) ;
\path[newedge] (n4) .. controls (5.090, -1.300) and (4.040, -1.300) ..  (n5) ;
\node[newlab] at (4.568, -1.301){src};
\path[quantedge] (n4)  -- node[lab]{at} (n7) ;
\path[deledge](n1.west |- 0.455, -1.855) -- node[dellab]{src} (n0) ;
\path[deledge](n4.west |- 3.525, -1.865) -- node[dellab]{trg} (n5) ;
\path[quantedge] (n0)  -- node[lab]{at} (n3) ;
\path[newedge] (n1) .. controls (2.020, -1.290) and (0.970, -1.290) ..  (n0) ;
\node[newlab] at (1.498, -1.291){trg};
\userdefinedmacro
\end{tikzpicture}
\renewcommand{\userdefinedmacro}{\relax}\caption{Reverse edges rule}\figlabel{reverse}\

\end{figure}

This rule has two parts, the first part shown on the left-hand side of \figref{reverse}
replaces all edges with label \src with edges with label \trg. The second part, shown in the right-hand
side of \figref{reverse} replaces all edges with label \trg by edges with label \src. Note that
this rule also complies with dangling edges as the rule does not require an edge to have both \src and \trg.
Moreover, because of the universal quantifier the rule is applied at once, and all edges are reversed by one rule application.
Performing this task without the use of a universal quantifier would need an extra control mechanism to avoid applying the rule forever.

\subsection{Simple migrations}
\GROOVE allows multiple type graphs to be used. In the migration case we enable both the source and the
target type graphs.

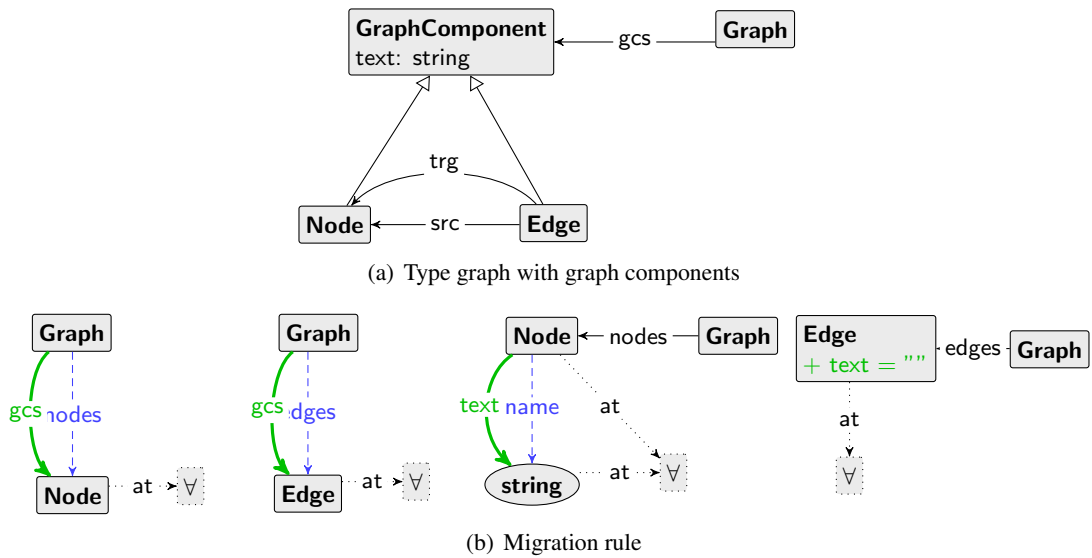
\begin{figure}
\vspace*{5mm}
\centering
\subfigure[Type graph with graph components]{
\scalebox{\mytikzscale}{
\scalebox{\mytikzscale}{\begin{tabular}[t]{@{}l@{}}
\begin{tikzpicture}[scale=\tikzscale]
\node[node] (n3)  at (2.870, -2.645) {\ml{\textbf{Edge}}};
\node[node] (n0)  at (4.210, -1.355) {\ml{\textbf{Graph}}};
\node[node] (n2)  at (1.415, -2.655) {\ml{\textbf{Node}}};
\node[node] (n1)  at (2.190, -1.440) {\ml{\textbf{GraphComponent}\\text: string}};
\path[edge](n0.west |- 2.190, -1.440) -- node[lab]{gcs} (n1) ;
\path[edge](n3.west |- 1.415, -2.655) -- node[lab]{src} (n2) ;
\path[edge] (n3) .. controls (2.500, -2.240) and (1.770, -2.250) ..  (n2) ;
\node[lab] at (2.142, -2.251){trg};
\path[subedge] (n3)  --  (n1) ;
\path[subedge] (n2)  --  (n1) ;
\userdefinedmacro
\end{tikzpicture}
\renewcommand{\userdefinedmacro}{\relax}\figlabel{graphcomponent}\end{tabular}}%

}
}
\subfigure[Migration rule]{
\scalebox{\mytikzscale}{
\scalebox{\mytikzscale}{\begin{tabular}[t]{@{}l@{}}
\begin{tikzpicture}[scale=\tikzscale]
\node[node] (n6)  at (3.765, -0.345) {\ml{\textbf{Node}}};
\node[node] (n3)  at (2.210, -1.395) {\ml{\textbf{Edge}}};
\node[node] (n0)  at (0.640, -0.325) {\ml{\textbf{Graph}}};
\node[node] (n7)  at (5.070, -0.345) {\ml{\textbf{Graph}}};
\node[node] (n9)  at (7.140, -0.435) {\ml{\textbf{Graph}}};
\node[quantnode] (n5)  at (2.930, -1.315) {\ml{$\forall$}};
\node[node] (n1)  at (0.645, -1.405) {\ml{\textbf{Node}}};
\node[node] (n4)  at (2.280, -0.325) {\ml{\textbf{Graph}}};
\node[quantnode] (n13)  at (5.810, -1.275) {\ml{$\forall$}};
\node[quantnode] (n2)  at (1.430, -1.345) {\ml{$\forall$}};
\node[node] (n11)  at (5.920, -0.430) {\ml{\textbf{Edge}\\{\color{\green}$+$ text = ""}}};
\node[quantnode] (n12)  at (4.640, -1.255) {\ml{$\forall$}};
\node[node, attr] (n8)  at (3.700, -1.335) {\ml{\textbf{string}}};
\path[newedge] (n6) .. controls (3.370, -0.600) and (3.320, -0.990) ..  (n8) ;
\node[newlab] at (3.352, -0.804){text};
\path[quantedge](n1.east |- 1.430, -1.345) -- node[lab]{at} (n2) ;
\path[quantedge](n11.south -| 5.810, -1.275) -- node[lab]{at} (n13) ;
\path[quantedge](n3.east |- 2.930, -1.315) -- node[lab]{at} (n5) ;
\path[quantedge] (n6)  -- node[lab]{at} (n12) ;
\path[edge](n9.west |- 5.920, -0.430) -- node[lab]{edges} (n11) ;
\path[deledge](n6.south -| 3.700, -1.335) -- node[dellab]{name} (n8) ;
\path[edge](n7.west |- 3.765, -0.345) -- node[lab]{nodes} (n6) ;
\path[deledge](n0.south -| 0.645, -1.405) -- node[dellab]{nodes} (n1) ;
\path[quantedge](n8.east |- 4.640, -1.255) -- node[lab]{at} (n12) ;
\path[deledge](n4.south -| 2.210, -1.395) -- node[dellab]{edges} (n3) ;
\path[newedge] (n4) .. controls (1.950, -0.570) and (1.920, -1.100) ..  (n3) ;
\node[newlab] at (1.940, -0.840){gcs};
\path[newedge] (n0) .. controls (0.320, -0.590) and (0.330, -1.130) ..  (n1) ;
\node[newlab] at (0.333, -0.865){gcs};
\userdefinedmacro
\end{tikzpicture}
\renewcommand{\userdefinedmacro}{\relax}\figlabel{tographcomponent}\end{tabular}}%

}
}
\caption{Type graph with the GraphComponent and its migration rule}
\end{figure}

\begin{description}
  \item[Graph component migration] The target type graph in \GROOVE shown in \figref{graphcomponent}
is notationally very similar to the one given in the case study description.
The migration rule shown in \figref{tographcomponent} consists of four parts, the first two parts in the left rename the labels \nodes and \edges to \gcs, the third part renames the ``name" attribute of nodes of type Node to ``text", and finally the fourth part of the rule adds an attribute ``text" to nodes of type Edge and initializes it with an empty string. All parts of the rule are universally quantified.

  \item[Topology changing migration] The type graph for the topology-changing migration case is shown in \figref{graphnoedge}. The migration rule is depicted in \figref{removeedges}. This rule has also three parts, the first part adds a new edge with the label \emph{linksTo} between any two nodes of type Node and removes the node Edge. The other two parts of the rule are to remove the dangling edges. Similar to the previous migration, all
      parts of the rule are universally quantified.
\end{description}

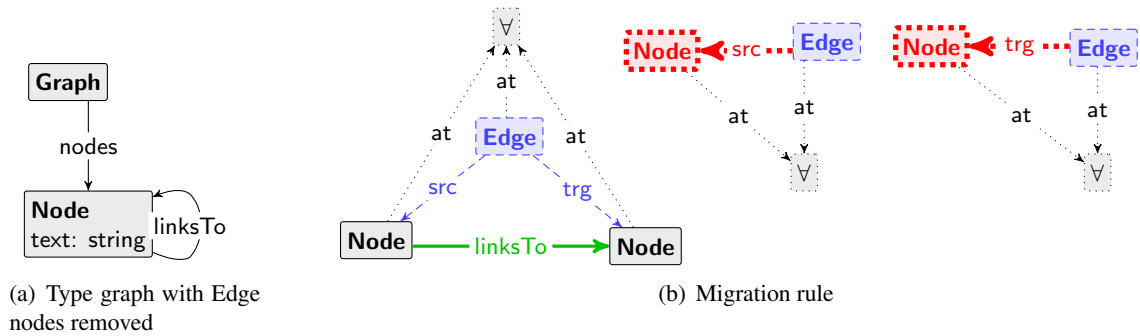
\begin{figure}
\vspace*{5mm}
\centering
\subfigure[Type graph with Edge nodes removed]{
\scalebox{\mytikzscale}{
\scalebox{\mytikzscale}{\begin{tabular}[t]{@{}l@{}}
\begin{tikzpicture}[scale=\tikzscale]
\node[node] (n0)  at (2.600, -1.345) {\ml{\textbf{Graph}}};
\node[node] (n1)  at (2.735, -2.290) {\ml{\textbf{Node}\\text: string}};
\path[edge](n0.south -| 2.735, -2.290) -- node[lab]{nodes} (n1) ;
\path[edge] (n1) .. controls (3.350, -2.560) and (3.410, -2.510) .. (3.410, -2.510).. controls (3.510, -2.400) and (3.510, -2.190) .. (3.410, -2.090).. controls (3.410, -2.090) and (3.350, -2.030) ..  (n1) ;
\node[lab] at (3.410, -2.301){linksTo};
\userdefinedmacro
\end{tikzpicture}
\renewcommand{\userdefinedmacro}{\relax}\figlabel{graphnoedge}\end{tabular}}%

}
}
\qquad
\subfigure[Migration rule]{
\scalebox{\mytikzscale}{
\scalebox{\mytikzscale}{\begin{tabular}[t]{@{}l@{}}
\begin{tikzpicture}[scale=\tikzscale]
\node[quantnode] (n6)  at (3.980, -2.505) {\ml{$\forall$}};
\node[node] (n3)  at (2.925, -2.995) {\ml{\textbf{Node}}};
\node[nacnode] (n0)  at (3.035, -1.695) {\ml{\textbf{Node}}};
\node[delnode] (n7)  at (5.970, -1.675) {\ml{\textbf{Edge}}};
\node[delnode] (n2)  at (2.020, -2.265) {\ml{\textbf{Edge}}};
\node[node] (n1)  at (1.135, -2.955) {\ml{\textbf{Node}}};
\node[delnode] (n5)  at (4.130, -1.635) {\ml{\textbf{Edge}}};
\node[nacnode] (n9)  at (4.825, -1.665) {\ml{\textbf{Node}}};
\node[quantnode] (n8)  at (5.930, -2.505) {\ml{$\forall$}};
\node[quantnode] (n4)  at (2.000, -1.535) {\ml{$\forall$}};
\path[quantedge] (n3)  -- node[lab]{at} (n4) ;
\path[newedge](n1.east |- 2.925, -2.995) -- node[newlab]{linksTo} (n3) ;
\path[nacedge](n5.west |- 3.035, -1.695) -- node[naclab]{src} (n0) ;
\path[quantedge](n7.south -| 5.930, -2.505) -- node[lab]{at} (n8) ;
\path[quantedge] (n9)  -- node[lab]{at} (n8) ;
\path[deledge] (n2)  -- node[dellab]{src} (n1) ;
\path[quantedge](n2.north -| 2.000, -1.535) -- node[lab]{at} (n4) ;
\path[quantedge] (n0)  -- node[lab]{at} (n6) ;
\path[quantedge] (n1)  -- node[lab]{at} (n4) ;
\path[nacedge](n7.west |- 4.825, -1.665) -- node[naclab]{trg} (n9) ;
\path[deledge] (n2)  -- node[dellab]{trg} (n3) ;
\path[quantedge](n5.south -| 3.980, -2.505) -- node[lab]{at} (n6) ;
\userdefinedmacro
\end{tikzpicture}
\renewcommand{\userdefinedmacro}{\relax}\figlabel{removeedges}\end{tabular}}%

}
}
\caption{Type graph and the migration rule for removing the nodified edges}
\end{figure}

\subsection{Delete node with specific name}
\begin{figure}
\centering
\subfigure[Delete node with name ``n1"]{
\scalebox{\mytikzscale}{
\scalebox{\mytikzscale}{\begin{tabular}[t]{@{}l@{}}
\begin{tikzpicture}[scale=\tikzscale]
\node[delnode] (n0)  at (1.065, 0.700) {\ml{\textbf{Node}\\name = "n1"}};
\node (n1) [below= of n0] {};
\userdefinedmacro
\end{tikzpicture}
\renewcommand{\userdefinedmacro}{\relax}\figlabel{deletenode}\end{tabular}}%

}
}
\subfigure[Delete node with name ``n1" and its incident edges]{
\scalebox{\mytikzscale}{
\scalebox{\mytikzscale}{\begin{tabular}[t]{@{}l@{}}
\begin{tikzpicture}[scale=\tikzscale]
\node[quantnode] (n3)  at (2.700, -1.825) {\ml{$\forall$}};
\node[delnode] (n0)  at (0.725, -1.150) {\ml{\textbf{Node}\\name = "n1"}};
\node[delnode] (n2)  at (1.700, -0.425) {\ml{\textbf{Edge}}};
\node[delnode] (n5)  at (1.850, -1.855) {\ml{\textbf{Edge}}};
\node[quantnode] (n4)  at (2.680, -0.395) {\ml{$\forall$}};
\path[quantedge](n5.east |- 2.700, -1.825) -- node[lab]{at} (n3) ;
\path[deledge] (n2)  -- node[dellab]{src} (n0) ;
\path[deledge] (n5)  -- node[dellab]{trg} (n0) ;
\path[quantedge](n2.east |- 2.680, -0.395) -- node[lab]{at} (n4) ;
\userdefinedmacro
\end{tikzpicture}
\renewcommand{\userdefinedmacro}{\relax}\figlabel{deletenodeandedges}\end{tabular}}%

}
}
\caption{Type graph and the migration rule}
\end{figure}
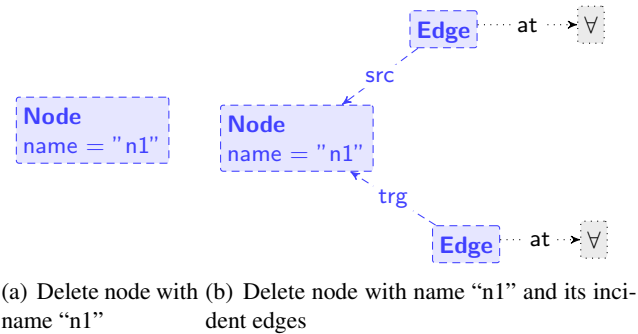

\begin{description}
  \item [Node deletion] Deleting a node with a specific name can be easily done in \GROOVE.
  In this case, we only need to have an eraser node with attribute name ``n1".
  Connected edges with labels \src and \trg are automatically deleted as \GROOVE uses
  single push-out rewriting (The rule is shown in \figref{deletenode}).
  \item [Node and incident edges deletion] In this case, nodes of type Edge which are connected to the node with name ``n1" must be explicitly deleted. Such nodes can be deleted using two separate universal quantifiers, one for  the edge nodes connected with an edge \src and one for edge nodes connected with an edge \trg (see \figref{deletenodeandedges}).
\end{description}

\subsection{Insert transitive edges}

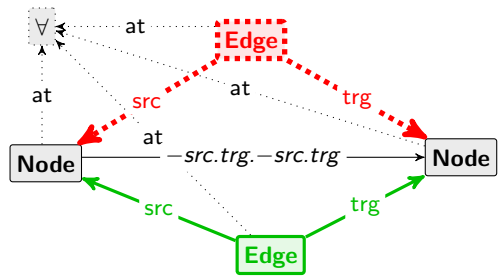
\begin{figure}
\vspace*{5mm}
\centering
\begin{tikzpicture}[scale=\tikzscale]
\node[newnode] (n3)  at (3.890, -2.945) {\ml{\textbf{Edge}}};
\node[node] (n0)  at (2.395, -2.335) {\ml{\textbf{Node}}};
\node[quantnode] (n2)  at (2.370, -1.425) {\ml{$\forall$}};
\node[node] (n1)  at (5.155, -2.295) {\ml{\textbf{Node}}};
\node[nacnode] (n4)  at (3.760, -1.515) {\ml{\textbf{Edge}}};
\path[quantedge] (n3)  -- node[lab]{at} (n2) ;
\path[newedge] (n3)  -- node[newlab]{trg} (n1) ;
\path[quantedge](n0.north -| 2.370, -1.425) -- node[lab]{at} (n2) ;
\path[quantedge] (n1)  -- node[lab]{at} (n2) ;
\path[nacedge] (n4)  -- node[naclab]{src} (n0) ;
\path[edge](n0.east |- 5.155, -2.295) -- node[lab]{\textit{$-$src.trg.$-$src.trg}} (n1) ;
\path[quantedge](n4.west |- 2.370, -1.425) -- node[lab]{at} (n2) ;
\path[newedge] (n3)  -- node[newlab]{src} (n0) ;
\path[nacedge] (n4)  -- node[naclab]{trg} (n1) ;
\userdefinedmacro
\end{tikzpicture}
\renewcommand{\userdefinedmacro}{\relax}\caption{Transitive closure rule}\figlabel{transitiveclosure}\

\end{figure}

The rule for inserting transitive edges is shown in \figref{transitiveclosure}. The rule checks
for the existence of a path with length two (two edge nodes) and the lack of a path of length one (one edge node).
The path with length two is specified using the regular expression ``-\src.\trg.-\src.\trg", as the identity of the intermediate edge nodes
is not important. Please note that this is just an abbreviation, we can instead specify a path
by explicitly specifying the edge nodes. The absence of a path of length one is specified by a NAC.
The insertion of a new edge is shown by a creator edge node and two edges.
Again all elements must be universally quantified as we want to insert transitive edges for the whole graph.

%
\section{Conclusion}
\seclabel{conclusion}
In this report we presented a \GROOVE solution to the Hello World case. We showed
that all requested operations including the optional ones can be solved easily. Each task
is solved using only one rule application of a single rule. All rules look very simple and
contain few nodes only. No control language or any other control mechanism was used
and all solutions solely use graph transformation mechanisms of \GROOVE. The grammar for the solution
can be found in the SHARE image \cite{share}.

\bibliographystyle{eptcs}
\bibliography{bib}


\end{document}